\documentclass{article}
\usepackage{spconf,amsmath,graphicx}
\usepackage{comment,psfrag,amssymb,booktabs,tabularx,pgfplots} 

\title{A Data Set Providing Synthetic and Real-World Fisheye Video Sequences}
\name{Andrea Eichenseer and Andr\'e Kaup}
\address{Multimedia Communications and Signal Processing\\
	Friedrich-Alexander University Erlangen-N\"urnberg (FAU), Cauerstr. 7, 91058 Erlangen, Germany\\
	}
\begin{document}
\ninept
\maketitle
\begin{abstract}
In video surveillance as well as automotive applications, so-called fisheye cameras are often employed to capture a very wide angle of view.
As such cameras depend on projections quite different from the classical perspective projection, the resulting fisheye image and video data correspondingly exhibits non-rectilinear image characteristics.
Typical image and video processing algorithms, however, are not designed for these fisheye characteristics.
To be able to develop and evaluate algorithms specifically adapted to fisheye images and videos, a corresponding test data set is therefore introduced in this paper.
The first of those sequences were generated during the authors' own work on motion estimation for fisheye videos and further sequences have gradually been added to create a more extensive collection.
The data set now comprises synthetically generated fisheye sequences, ranging from simple patterns to more complex scenes, as well as fisheye video sequences captured with an actual fisheye camera.
For the synthetic sequences, exact information on the lens employed is available, thus facilitating both verification and evaluation of any adapted algorithms.
For the real-world sequences, we provide calibration data as well as the settings used during acquisition.
The sequences are freely available via \texttt{www.lms.lnt.de/fisheyedataset/}.
\end{abstract}
\begin{keywords}
Data Set, Fisheye Video Sequences, Test Images, Synthetic Sequences, Sensor Data
\end{keywords}
\section{Introduction}
\label{sec:intro}

Automotive applications like advanced driver's assistance systems often require very large fields of view (FOV) to provide the necessary information about the car's environment.
In video surveillance, where it is also fundamental to survey large areas, an FOV of well beyond the common 40 to 60 degrees typical lenses provide is desirable so that as few cameras as possible may be installed. 
Fisheye lenses~\cite{miyamoto1964fel} are capable of achieving an FOV of 180 degrees and more and are thus very much suited to the aforementioned application scenarios.
In automotive applications, fisheye cameras are made use of for lane detection~\cite{auto1} and similar assistance systems~\cite{gehrig2, gehrig}; in surveillance scenarios, object tracking~\cite{surveillance2} and the generation of perspective views~\cite{surveillance} are important aspects.

A typical characteristic of fisheye imagery is the strong radial distortion that is introduced by projecting a hemisphere onto the image plane.
This characteristic is not taken into account by conventional image and video processing algorithms, which may perform poorly when being applied to fisheye data.
As fisheye cameras gain more and more popularity, and are now used also by outdoor enthusiasts (the wide-angle \textit{GoPro} camera is just one example), the need for algorithms adapted to fisheye data thus arises. 

\begin{figure}[t]
\small
\label{fig:projections}
\centering
\psfrag{e}[cc][cc]{\small \hspace{0.15cm}Equisolid fisheye, FOV~=~185$^\circ$}
\psfrag{o}[cc][cc]{\small Perspective, FOV~$\approx$~110$^\circ$}
\centerline{\includegraphics[width=0.9\columnwidth]{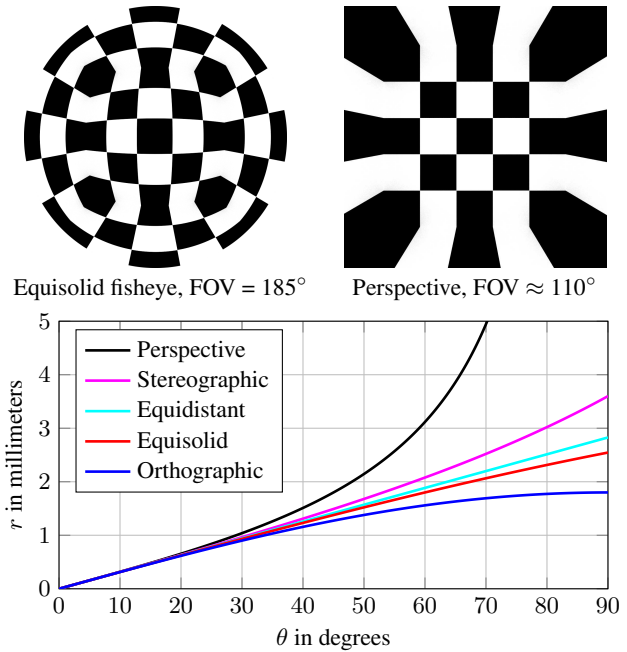}}
\input{models}
\vspace{-0.4cm}
\caption{Top: Comparison of equisolid fisheye FOV and perspective FOV (same sensor size). Bottom: Projections as functions of radius $r$ over incident angle $\theta$ with $f$ = 1.8 mm and $\theta_{\text{max}}$ = 90$^\circ$.}
\vspace{-0.5cm}
\end{figure}

During the development of new image processing algorithms, test data such as the well-known Kodak image set~\cite{kodak} or the Tecnick test images~\cite{tecnick, tecnick2} are frequently used.
For multiview applications, the Middlebury set~\cite{middlebury} is available and for video processing and coding, one might make use of the video set provided by ARRI~\cite{arri} and the HEVC Class A through F test video sequences~\cite{hevctest}, respectively.
When dealing with non-rectilinear imagery, however, one is hard pressed to find freely available data sets similar to those mentioned.
To the best of the authors' knowledge, no such data set exists for fisheye video sequences, either.

Because of this apparent lack of suitable fisheye material, a data set first created by the authors during their previous work on motion estimation and temporal error concealment for fisheye video sequences~\cite{eichenseer2015motionfish, eichenseer2015tecfish}, and significantly extended since, is introduced in this paper and made freely available via their website.
Both synthetically generated and actually captured fisheye video sequences are provided, ranging from simple patterns to more complex scenes, including automotive and surveillance content.
For all sequences, information on the acquisition process as well as further parameters are provided.
Making use of the synthetic sequences especially facilitates the verification and evaluation of new algorithms, since here, all parameters are known and an exact projection function is used so that calibration is unnecessary. 
To demonstrate the use of the data set and to show a practical application for our sequences, we revisit our previously published hybrid motion estimation method for fisheye video sequences~\cite{eichenseer2015motionfish} and provide some new results for the synthetic sequences at the end of the paper.

\begin{table}[t]
\small
\caption{Perspective projection and classical fisheye projections.}
\label{tab:projections}
\centering
\renewcommand\arraystretch{0.95}
\begin{tabularx}{\columnwidth}{Xlcc}
\toprule
Projection & Equation & FOV$_{\text{max}}$ & $\theta_{\text{max}}$ \\
\midrule
Perspective & $r = f \tan \theta$ & $\ll$ 180$^{\circ}$ & $\ll$ 90$^{\circ}$ \\
\addlinespace
Equidistant & $r = f\theta$ & $\infty$ & $\infty$ \\
Equisolid & $r = 2f \sin{\left( \theta/2\right)}$ & 360$^{\circ}$ & 180$^{\circ}$ \\
Stereographic & $r = 2f \tan{\left(\theta/2\right)}$ & $\ll$ 360$^{\circ}$ & $\ll$ 180$^{\circ}$\\
Orthographic & $r = f \sin{\theta}$ & 180$^{\circ}$ & 90$^{\circ}$ \\
\bottomrule
\end{tabularx}
\vspace{-0.5cm}
\end{table}

\section{Synthetic Fisheye Video Sequences}
\label{sec:projections}

\begin{table}[t]
\small
\caption{Global \textit{Blender} configuration.}
\label{tab:blender}
\centering
\renewcommand\arraystretch{0.95}
\begin{tabularx}{\columnwidth}{Xl}
\toprule
Engine & Cycles Renderer \\
Lens & Panoramic, Fisheye Equisolid (Equidistant)\\
Focal length & 1.8 mm (not needed for equidistant model) \\
Field of view & 185 degrees \\
Sensor size & 5.20 mm, AUTO (5.81 mm, AUTO) \\
Resolution & 1088 $\times$ 1088 pixels \\
Output & PNG, RGBA, 8 bits, uncompressed \\
Noise pattern & Varying for each frame \\ 
\bottomrule
\end{tabularx}
\vspace{-0.6cm}
\end{table}
Images and videos captured by conventional cameras follow the pinhole model, i.\,e., a perspective projection is used to map a scene to the image plane.
In contrast to that, fisheye lenses make use of projections which are able to capture a far wider FOV~\cite{miyamoto1964fel,kannala2006genericmodel}. The four classical fisheye projections as well as the perspective projection are compared in Table~\ref{tab:projections}, with $f$, $\theta$, and $r$ denoting the focal length, incident angle of light, and distance to the image center, respectively.
The maximum theoretically possible FOV as well as the corresponding maximum $\theta$ are also given.
Fig.~\ref{fig:projections} visualizes the different projection functions.
Here, the limitations of the pinhole model's FOV become very evident when the distance $r$ is actually interpreted as half the sensor width.
Quite obviously, a large FOV can only be obtained by making use of a fisheye projection.

The classical fisheye projections are given by trigonometrical expressions which can be easily inverted.
To inspect basic characteristics of fisheye images and their behavior in typical image processing algorithms, we thus decided to create synthetic fisheye image sequences so as to have control over the exact underlying model.
As a rendering software, \textit{Blender}~\cite{blender} was employed as it supports perspective as well as panoramic lenses.
For the latter, both equisolid and equidistant fisheye projections are available, which we made use of for our synthetic sequences.
The global settings for the equisolid sequences are summarized in Table~\ref{tab:blender} with differing settings for the equidistant versions denoted in parentheses.
The choice of $f$ = 1.8 mm is due to the actual fisheye lens used for the recording of the real-world sequences.
In order to generate circular equisolid fisheye images, we derived a virtual sensor of size 5.20 by 5.20 mm as this just allows capturing an FOV of 185 degrees when employing the equisolid model: $2r = 2f\sin{(\theta_{\text{max}}/2)} \approx 5.20$, with $\theta_{\text{max}} = \text{FOV}/2$ and $2r$ representing the corresponding sensor width.
For the equidistant model, we require $2r = f\theta_{\text{max}} \approx 5.81$, i.\,e., a sensor size of 5.81 by 5.81 mm.

\begin{figure*}[t]
\centering
\psfrag{a1}[lt][lt]{\small A2}
\psfrag{a2}[lt][lt]{\small A3}
\psfrag{a3}[lt][lt]{\small H2}
\psfrag{a4}[lt][lt]{\small F1}
\psfrag{a5}[lt][lt]{\small F3}
\psfrag{a6}[lt][lt]{\small B1}
\psfrag{a7}[lt][lt]{\small E2}
\psfrag{a8}[lt][lt]{\small G1}
\psfrag{a9}[lt][lt]{\small G4}
\psfrag{a0}[lt][lt]{\small K1}
\psfrag{o1}[lt][lt]{\small C2}
\psfrag{o2}[lt][lt]{\small E6}
\psfrag{o3}[lt][lt]{\small J1}
\psfrag{o4}[lt][lt]{\small J3}
\psfrag{o5}[lt][lt]{\small J4}
\centerline{\includegraphics[width=0.98\textwidth]{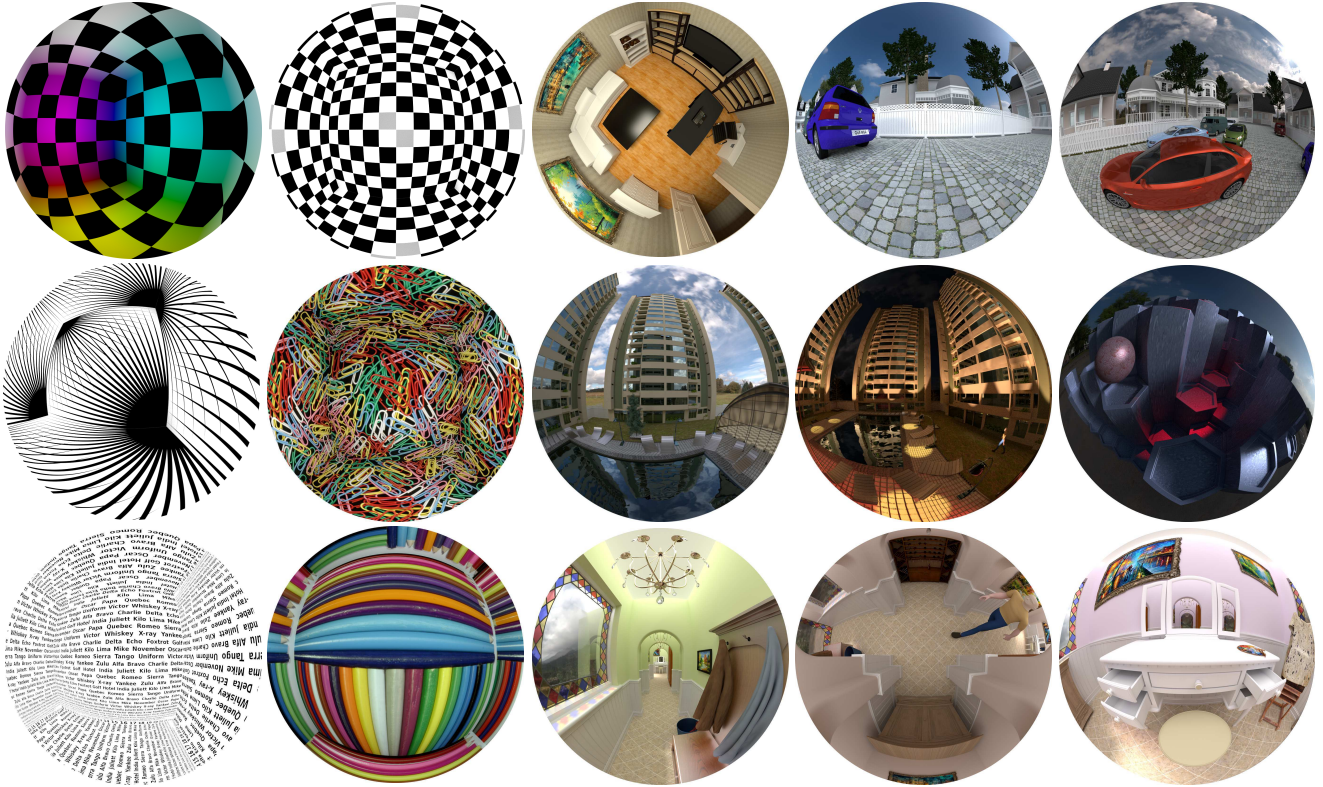}}
\vspace{-0.3cm}
\caption{Example images of some of the synthetic equisolid fisheye sequences.}
\label{fig:synthSeq}
\vspace{-0.5cm}
\end{figure*}

Table~\ref{tab:charsynth} provides the characteristics of the synthetic sequences.
Fig.~\ref{fig:synthSeq} shows example frames of the majority of those sequences, with synthetic pattern sequences on the left and synthetic scene sequences on the right side.
The former make use of simple patterns such as checkerboards arranged on a cube around a moving fisheye camera, which is identical for sequences A1 through E6 and provides translational motion as well as a horizontal pan, zoom, and vertical pan.
\textit{CheckercubeC}, \textit{CheckercubeD}, \textit{Rays}, \textit{Stars}, and \textit{Lorem} are based on images taken from the artificial image set of the Tecnick~\cite{tecnick} library.
\textit{CheckercubeC} can also be made use of for calibration purposes.
As the underlying model is mathematically exact, it can be used to verify calibration methods based on checkerboard patterns such as~\cite{scara2}.
\textit{Cards}, \textit{Clips}, \textit{Coins}, \textit{Fence}, \textit{Flowers}, and \textit{Pencils} use images taken from the new Tecnick~\cite{tecnick2} image set.

For the synthetic scene sequences, we combined several object models available through~\cite{blendswap} to create suitable scenes.
We then added the fisheye camera, set both camera and object motion as desired, and thus generated more realistic sequences.
The motion between two consecutive frames was generally chosen to be minor.
Therefore, the frame rate should be set to 50 Hz during playback.
Depending on user preference, a lower frame rate (starting from 25 Hz) may also be reasonable.
If needed, the sequences can be temporally subsampled to create more pronounced motion patterns. 

\begin{table}[t]
\small
\caption{Characteristics of the synthetic sequences.}
\label{tab:charsynth}
\centering
\renewcommand\arraystretch{0.92}
\begin{tabularx}{\columnwidth}{lXrrll}
\toprule
ID & Name & Frames & Camera & Scene & Motion \\
\midrule
A1 & CheckercubeA & 350 & moving & static & TPZ \\
A2 & CheckercubeB & 350 & moving & static & TPZ \\
A3 & CheckercubeC & 350 & moving & static & TPZ \\
A4 & CheckercubeD & 350 & moving & static & TPZ \\
B1 & Rays & 350 & moving& static & TPZ \\
B2 & Stars & 350 & moving& static & TPZ \\
C1 & Lorem & 350 & moving& static & TPZ \\
C2 & Alphabet & 350 & moving& static & TPZ \\
D1 & Gradient & 350 & moving& static & TPZ \\
E1 & Cards & 350 & moving& static & TPZ \\
E2 & Clips & 350 & moving& static & TPZ \\
E3 & Coins & 350 & moving& static & TPZ \\
E4 & Fence & 350 & moving & static & TPZ \\
E5 & Flowers & 350 & moving & static & TPZ \\
E6 & Pencils & 350 & moving& static & TPZ \\
\addlinespace
F1 & Street & 1800 & moving& static & TZ \\
F2 & CarsA & 121 & static& moving & T \\
F3 & CarsB & 121 & static& moving & T \\
G1 & PoolA & 752 & moving& static & T \\
G2 & PoolB & 748 & moving& static & T \\
G3 & PoolNightA & 320 & moving& moving & TP \\
G4 & PoolNightB & 320 & static& moving & T \\
H1 & Room & 400 & static& moving & T \\
H2 & LivingroomA & 111 & moving & static & R \\
H3 & LivingroomB & 82 & static & moving & T \\
H4 & LivingroomC & 82 & moving & static & T\\
J1 & HallwayA & 600 & moving& static & ZP \\
J2 & HallwayB & 120 & moving& moving & RT \\
J3 & HallwayC & 120 & static & moving & T \\
J4 & HallwayD & 500 & moving & static & TP \\
K1 & PillarsA & 550 & static& moving & T \\
K2 & PillarsB & 550 & static& moving & T \\
K3 & PillarsC & 550 & moving& moving & T \\
\midrule
\multicolumn{6}{r}{\footnotesize T: Translation, Z: Zoom, P: Pan, R: Rotation} \\
\bottomrule
\end{tabularx}
\vspace{-0.6cm}
\end{table}

\section{Real-World Fisheye Video Sequences}

\begin{figure*}[t]
\centering
\psfrag{a1}[cc][cc]{\small TestchartA}
\psfrag{a2}[cc][cc]{\small LabRoomA}
\psfrag{a3}[cc][cc]{\small LibraryA}
\psfrag{a4}[cc][cc]{\small LibraryB}
\psfrag{a5}[cc][cc]{\small ClutterA}
\psfrag{a6}[cc][cc]{\small LectureA}
\psfrag{a7}[cc][cc]{\small CarparkC}
\psfrag{a8}[cc][cc]{\small CarparkB}
\psfrag{a9}[cc][cc]{\small ElevatorA}
\psfrag{a0}[cc][cc]{\small DriveA}
\centerline{\includegraphics[width=\textwidth]{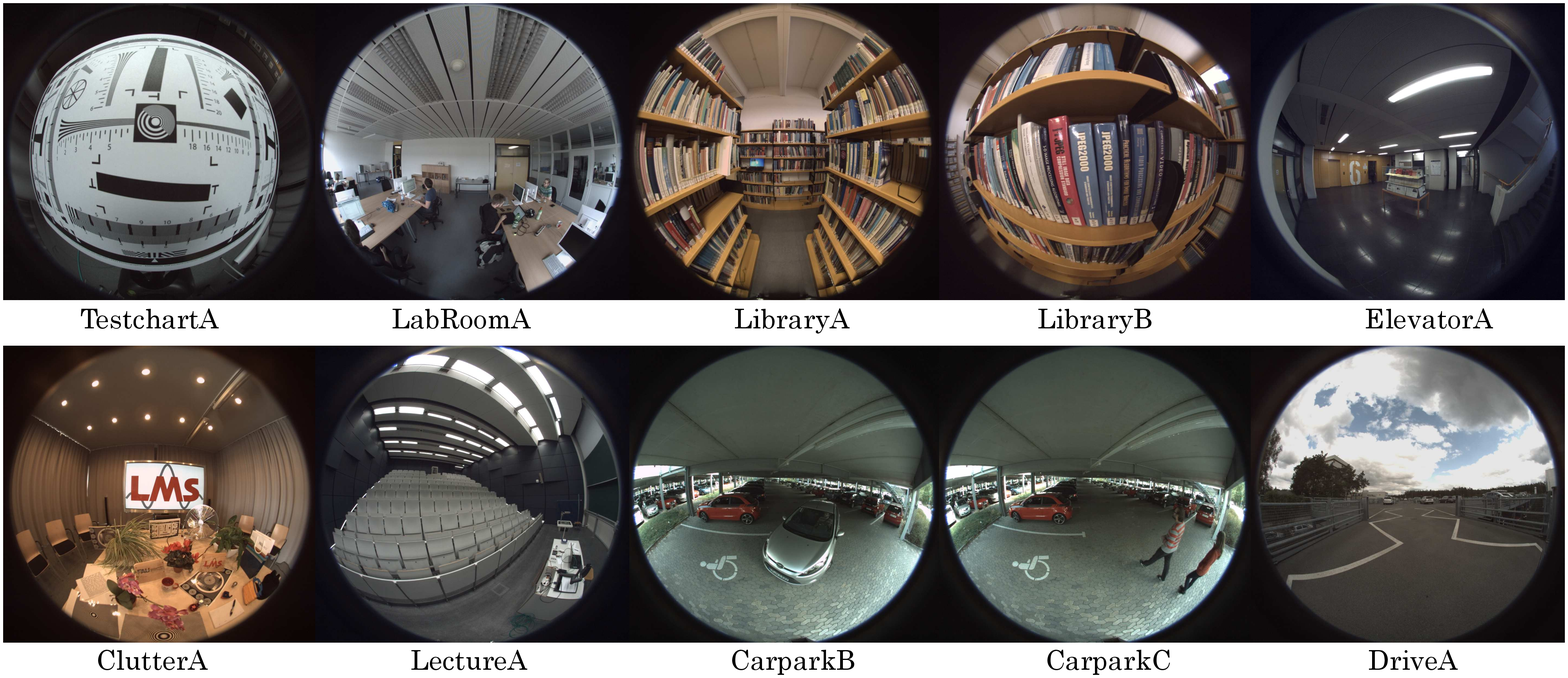}}
\vspace{-0.4cm}
\caption{Example frames of some of the real-world sequences.}
\label{fig:realScene}
\vspace{-0.7cm}
\end{figure*}

Apart from synthetic sequences, the data set also contains a number of real-world sequences that were captured with an actual fisheye camera.
The camera employed was a Basler ace acA2000-50gc which uses a GigE interface for recording up to 50 frames per second with a resolution of 2048$\times$1086 pixels.
The raw output images are saved in 12 bit Bayer GR format.
As a lens, the Fujinon FE185C057HA-1 fisheye was used.
This fixed-focus lens has a focal length of 1.8 mm, a manually operated iris range of F1.4$\sim$F16, and allows capturing an FOV of 185 degrees.
As the camera contains a 2/3~inch sensor, the lens thus creates circular fisheye images.


For our data set, we chose mainly indoor scenes, including surveillance scenarios, as the large FOV easily leads to overexposed areas which are difficult to control.
An advantage of indoor locations is given by the many straight lines that can be found on ceilings, furniture, etc., and as straight lines are indeed not straight when captured with a fisheye camera, they serve as the prime characteristic in almost every sequence.
Despite the difficulty of exposure control, some outdoor sequences were recorded representing automotive and surveillance scenarios.
Both exposure time and aperture were adjusted to the respective lighting conditions and may be taken from Table~\ref{tab:charreal}, where the characteristics of the real-world sequences can be found. 
The target frame rate was set to a constant 40 frames per second, but may actually be less than that.
Since the circular fisheye images occupy a roughly square sensor area, all images recorded have a cropped resolution of 1150~$\times$1086~pixels and may be further cropped as needed.
All sequences are provided in both raw image 12 bit TIFF format (sensor data) and in 8 bit PNG format (output data).
For the latter, the AHD~\cite{hirakawa2005ahd} algorithm was employed for demosaicking preceded by white balancing and followed by gamma correction in order to create a pleasing subjective impression of the scene.
Fig.~\ref{fig:realScene} provides example frames for many of the real-world sequences.

Since actual fisheye lenses rarely follow one of the models described in section~\ref{sec:projections}, only a calibration helps to determine the camera parameters.
We thus also provide calibration images in our data set, which can be used with calibration methods that are based on a calibration pattern.
The provided images show a checkerboard pattern at different angles and positions and have been successfully tested for calibration via the OcamCalib Toolbox version 3.0~\cite{scara2}.
The calibration results are made available along with the calibration images.

\section{Fisheye Motion Estimation}

As initially motivated, conventional video processing algorithms do not take into account the fisheye characteristics.
Traditional block-matching motion estimation (ME)~\cite{blockmatching}, for example, is based on a translational motion model as this is the assumed predominant kind of motion found in typical video sequences.
Since this kind of ME method works in a block-based manner, it is suited to global and local translational motion alike.
With fisheye video sequences, however, the translational motion model is no longer valid.
In~\cite{eichenseer2015motionfish}, we therefore introduced a hybrid ME technique adapted to equisolid fisheye data which introduces suitable projections based on the equisolid fisheye model so as to be able to exploit the translational model while at the same time taking the fisheye characteristics into consideration.
A small subset of the synthetic sequences presented here has already been used for evaluating this method.
We would now like to provide some additional results and thereby show that it is indeed necessary to adapt conventional image and video processing techniques to fisheye data.
To that end, Table~\ref{tab:psnr} provides a comparison of traditional, translational motion estimation (TME) with our hybrid technique (HME).
The search range and block size employed here were 64 and 16 pixels, respectively.
Details about the algorithm as well as further settings may be taken directly from~\cite{eichenseer2015motionfish}. 
With average obtainable gains of up to 2 dB, the need for fisheye adaptations is clearly given and thus, our proposed data set aims at helping develop these adaptations.

\begin{table}[h!]
\small
\caption{Characteristics of the real-world sequences.}
\label{tab:charreal}
\centering
\renewcommand\arraystretch{0.95}
\begin{tabularx}{\columnwidth}{Xrccl}
\toprule
Name & Frames & Camera/Scene: & Exposure & Aper- \\
& & Motion & Time & ture\\
\midrule
TestchartA & 300 & static/moving: T & 10 ms & F4\\
TestchartB & 300 & static/moving: T & 10 ms & F4\\
TestchartC & 500 & static/shaky: T & 10 ms & F4 \\
AlfaA & 500 & static/moving: T & 10 ms & F4\\
AlfaB & 500 & static/moving: T & 10 ms & F4\\
AlfaC & 500 & static/shaky: T & 10 ms & F4 \\
LibraryA & 500 & moving/static: Z & 20 ms & F1.4 \\
LibraryB & 487 & moving/static: T & 20 ms & F1.4 \\
LibraryC & 252 & moving/static: P & 20 ms & F1.4\\
LibraryD & 300 & shaky/static: T & 20 ms & F1.4 \\
LibraryE & 300 & shaky/static: T & 20 ms & F1.4 \\
CarparkA & 302 & static/moving: T & 15 ms & F6\\
CarparkB & 982 & static/moving: T & 15 ms & F6\\
CarparkC & 1387 & static/moving: T & 15 ms & F6 \\
DriveA & 1251 & moving/static: ZP & 1 ms & F6\\
DriveB & 400 & moving/moving: ZPT & 1 ms & F6\\
DriveC & 400 & moving/static: ZP & 1 ms& F6\\
DriveD & 863 & moving/static: TP & 5 ms& F16\\
DriveE & 420 & moving/static: T & 5 ms& F16\\
ElevatorA & 321 & static/moving: T & 15 ms & F4 \\
ElevatorB & 211 & static/moving: T & 15 ms & F4 \\
ElevatorC & 136 & static/moving: T & 15 ms & F4\\
ElevatorD & 216 & static/moving: T & 15 ms & F4\\
LabRoomA & 650 & static/moving: T & 15 ms & F4 \\
LabRoomB & 241 & static/moving: T & 15 ms & F4 \\
ClutterA & 500 & moving/static: T & 30 ms & F1.4 \\
ClutterB & 500 & moving/static: T & 30 ms & F1.4 \\
LectureA & 846 & moving/static: P & 15 ms & F4\\
LectureB & 461 & moving/static: T & 15 ms & F4\\
\addlinespace
Calibration & -- & static/various angles & 15 ms & F4 \\
\midrule
\multicolumn{5}{r}{\footnotesize T: Translation, Z: Zoom, P: Pan} \\
\bottomrule
\end{tabularx}
\vspace{-0.3cm}
\end{table}

\vspace{-0.3cm}
\section{Conclusion}
\label{sec:conclusion}
\vspace{-0.15cm}

This paper introduced a data set of fisheye video sequences that may be used in the development of new image and video processing algorithms designed for fisheye imagery.
While one part of the data set comprises real-world sequences captured with an actual fisheye camera, including material for automotive and video surveillance scenarios, the other part of the set provides synthetic fisheye sequences generated in \textit{Blender}.
Due to their mathematical exactness, the synthetic sequences are especially useful as ground truth data on which novel algorithms may initially be tested.
Parameters like focal length and resolution were chosen such that the synthetic sequence characteristics are very similar to the real-world sequence characteristics.
Any algorithms should thus be very well applicable to both categories of sequences.
For the real-world sequences, calibration images and results are provided so as to allow adaptation to more realistic camera parameters. 
The need for fisheye adaptations was demonstrated with a brief excursion into fisheye motion estimation.
Further sequences and ground-truth motion information for the synthetic sequences are currently being worked on and will be progressively made available via the authors' website \texttt{www.lms.lnt.de/fisheyedataset/}.

\begin{table}[t]
\small
\caption{Average luminance PSNR results and gain per sequence.}
\label{tab:psnr}
\centering
\renewcommand\arraystretch{1.0}
\begin{tabularx}{\columnwidth}{lccc}
\toprule
Sequence (frames tested) & HME & TME & Gain \\
\midrule
CheckercubeC (34) & 32.62 dB & 30.59 dB & 2.03 dB \\
Alphabet (34) & 21.33 dB & 19.58 dB & 1.76 dB \\
Rays (34) & 28.08 dB & 26.77 dB & 1.31 dB \\
PoolA (74) & 39.76 dB & 37.85 dB  & 1.91 dB \\
PoolNightA (63) & 35.80 dB& 34.40 dB & 1.40 dB \\
PillarsA (109) & 42.42 dB & 40.74 dB & 1.68 dB \\
\bottomrule
\end{tabularx}
\vspace{-0.4cm}
\end{table}

\section{Acknowledgment}

The authors would like to thank Michael Ortner and Andreas Luder for their work on the synthetic and real-world sequences, respectively.
Detailed credits on the Blender models used may be found on the authors' webpage.
This work was supported by the Research Training Group 1773 “Heterogeneous Image Systems”, funded by the German Research Foundation (DFG).


\bibliographystyle{IEEEbib}
\bibliography{strings,refs}

\end{document}